\documentclass[12pt,preprint]{aastex}

\shorttitle{Promiscuous Stars}
\shortauthors{Hurley et al.}

\slugcomment{To appear in the Astrophysical Journal}

\begin{document}

\title{The Promiscuous Nature of Stars in Clusters}

\author{Jarrod R. Hurley and Michael M. Shara}
\affil{Department of Astrophysics, 
       American Museum of Natural History, \\ 
       Central Park West at 79th Street, 
       New York, NY 10024}
\email{jhurley@amnh.org, mshara@amnh.org}

\begin{abstract}
The recent availability of special purpose computers designed 
for calculating gravitational interactions of $N$-bodies at extremely high 
speed has provided the means to model globular clusters on a star-by-star 
basis for the first time. 
By endeavouring to make the $N$-body codes that 
operate on these machines as realistic as possible, 
the addition of stellar evolution being one example, much is 
being learnt about the interaction between the star cluster itself 
and the stars it contains. 
A fascinating aspect of this research is the ability to follow the 
orbits of individual stars in detail and to document the formation of 
observed exotic systems. 
This has revealed that many stars within a star cluster lead wildly 
promiscuous lives, interacting, often intimately and in rapid succession, 
with a variety of neighbours. 
\end{abstract}


\section{Introduction}
\label{s:intro}

The rich environment of a star cluster provides an ideal laboratory for the 
study of self-gravitating systems. 
Star clusters within our Galaxy range in size from loose associations and 
open clusters, which contain up to tens of thousands of stars, to the large 
globular clusters which host a million stars or more. 
Globular clusters are some of the oldest objects known and dynamically 
they lie in a very interesting regime. 
Compared to the solar neighbourhood the central density of stars is high 
enough (a factor of 10 million or more greater in some globular clusters)  
that a significant fraction of the cluster stars are likely to experience 
at least one close encounter with another star during their lifetime. 
At the opposite end of the scale, the size of a globular cluster is small 
enough, in comparison to a galaxy, that its dynamical relaxation 
timescale is less than its age. 
These combined considerations make a globular cluster an exciting place 
for a star to reside.  
As such we expect the stellar populations of a star cluster, the various 
classes of stars and binaries, to exhibit a dynamical signature.  
In particular, many of the stars contained in the cluster will deviate 
strongly from the evolutionary paths predicted by standard stellar and 
binary evolution theory. 
This is indeed what we observe: either when pointing the 
Hubble Space Telescope (HST) at the centres of globular clusters or 
when generating star-by-star models of star cluster evolution.

\section{Observations of Exotic Objects}
\label{s:observe}

Observational evidence for the modification of stellar populations
in star clusters owing to dynamical interactions between the cluster
stars is found in many instances (see Bailyn 1995 for a review).
A common indicator of this effect is a variation in the radial
distribution of a particular stellar sub-population within the cluster.
If, for example, a certain type of exotic star is found primarily in
the core of a cluster, where the number density of stars is highest and
hence interactions between stars are most likely, this is taken as strong
evidence that these stars have a dynamical origin.

Blue stragglers (BSs) are stars burning hydrogen in their cores that,  
according to their stellar mass, should already have evolved off the 
main-sequence (MS) and past the red giant stage to become 
white dwarfs (WDs): stellar corpses. 
Somehow these stars have been rejuvenated. 
Popular theories of the cause of this phenomenon are mass transfer from 
a companion star in a close binary system or the collision of two 
MS stars. 
HST observations of the dynamically old and metal-poor Galactic globular 
cluster M30 have revealed a number of BSs 
(Guhathakurta et al. 1998). 
When plotted in a cumulative radial distribution these BSs appear 
concentrated towards the cluster core relative to standard MS stars. 
Conversely, the population of bright red giants in M30 
is depleted in the core. 
This suggests that collisions are destroying giants, by ripping off their 
envelopes and reducing them to WDs, and creating BSs. 
Ferraro et al. (1999) used HST to identify 305 BSs in the high density globular 
cluster M80. 
These are also centrally concentrated with respect to the giants in the cluster 
and the suggestion is that M80 is being observed at a critical phase in 
its evolution when stellar interactions are delaying the collapse of the core 
(Ferraro et al. 1999) - leading to the production of collisional BSs. 

Blue Stragglers are not the only stars exhibiting enhanced populations 
in the cores of globular clusters. 
X-ray binaries are at least $1\,000$ times more abundant in 
clusters than in the Galactic field, and are being found in 
ever-increasing numbers with the Chandra X-ray observatory 
(Grindlay et al. 2001). 
Millisecond pulsars, including some in binaries or with planets 
(D'Amico et al. 2001) are prevalent in cluster cores. 
Cataclysmic binaries in clusters have been predicted and are also being 
found (Shara et al. 1996; Grindlay et al. 2001). 
A few sdB stars have been located and characterized (Moehler et al. 1997). 

Even though the stellar neighbourhood within an open cluster is less dense 
than that of a globular cluster it is still capable of producing 
exotic objects. 
The number of BSs found in the open cluster M67 is much 
greater than we would expect if they are simply produced via 
mass transfer in binaries exhibiting the same distribution of orbital 
characteristics as binary stars found in the field. 
Furthermore, these BSs have a variety of living arrangements 
(Leonard 1996, and references within). 
Some are single while others are found with a companion. 
In some cases the BS and the companion star interact in an 
intimate and regular manner (short-period circular orbit), in other  
cases the relationship is distant (long-period orbit) and may even 
appear eccentric. 
One of the BSs is so massive, a super-BS, that it must 
represent the merger of three stars, and another is observed in an active 
triple-system, a menage-a-trois (van den Berg et al. 2001). 
All of this suggests that the BSs have varied formation 
histories and that no one mechanism is responsible for their 
production (Leonard 1996). 
In particular, the presence of a super-BS and of BSs 
in eccentric binaries is not predicted by standard binary evolution and 
points to dynamical interactions tampering with the destinies of 
stars.

\section{Simulation Methods}
\label{s:method}

On a basic dynamical level a star cluster is composed of $N$ bodies
interacting with each other due to the gravitational force of every other
body in the system.
The ideal method for following the evolution of such a system is to integrate
directly the $N$ individual equations of motion, the $N$-body approach.
However, the cost of integrating the cluster for a relaxation-time scales 
as $N^3$ -- $N$ per force calculation, $N^2$ to integrate $N$ bodies for one 
crossing (dynamical) time, and there are of order $N$ crossing times 
per relaxation time -- so the method is computationally expensive. 
As a result most $N$-body simulations performed until recently, 
although state-of-the-art at the time, 
have involved a varying number of simplified and unrealistic 
conditions, such as including only single stars, using only equal-mass 
stars, neglecting stellar evolution or assuming no external tidal field 
(McMillan, Hut \& Makino 1991; Heggie \& Aarseth 1992). 
Performance has been enhanced by the development of improved computational 
algorithms. 
For example, the use of individual time-steps (Aarseth 1963) enables 
each star to evolve on its own natural dynamical timescale rather 
rather than forcing a small time-step on the entire system when only 
a fraction of the stars may need frequent attention. 
Quantizing these time-steps into a series of hierarchical levels 
increases the efficiency of the integration (McMillan 1986). 
Also, regularization techniques have been developed to remove the singularities
involved with compact binaries and few-body configurations, as well as
to improve the accuracy of the treatment (Mikkola \& Aarseth 1998). 

The very nature of the large-$N$ effects which make the $N$-body method so
computationally expensive can be used to justify the use of statistical
algorithms such as the orbit-averaged Fokker-Planck equation 
(H\'{e}non 1961; Giersz 2001; Joshi, Nave, \& Rasio 2001). 
In theory the $N$-body and Fokker-Planck methods, when applied to the same
initial conditions, should give similar results in the limit of large $N$ 
(Takahashi \& Portegies Zwart 2000). 
Ultimately the $N$-body approach remains the method of choice for creating
dynamical models of star clusters as a minimum number of simplifying
assumptions are required and it is relatively easy to implement additional
realistic features (see Meylan \& Heggie 1997 for a full discussion). 
It is also possible to follow the journey made by a particular star as 
the cluster evolves. 

A major impact on the suitability of $N$-body methods has been recent
advances in computing power, in particular the development of special-purpose
hardware.
The GRAPE-4 system became commercially available in 1995 and was designed
specifically for high-accuracy simulations of dense stellar systems
(Makino \& Taiji 1998).
It basically acts as a Newtonian force accelerator: the host workstation
supplies the GRAPE (GRAvity PipE) with the positions, velocities and masses
of a list of particles and the GRAPE returns the force and its time derivative
for each particle.
A 96-chip GRAPE-4 board operates at a peak speed of $\sim 50\,$Gflops 
(50 billion floating-point operations per second) 
and has enabled production of realistic open cluster models in a 
reasonable time.
To be precise, a simulation of $10\,000$ stars with a moderate binary
fraction of 10\% could be completed within a week. 
Dealing with binary systems complicates matters because as far as the 
GRAPE is concerned it sees binaries solely as the centre-of-mass particle 
and the integration of the orbits of the stars within each binary system 
must be dealt with on the host machine. 
Therefore the inclusion of a large proportion of binaries in a
simulation is a potential bottle-neck to the computational performance. 
These concerns are non-trivial owing to the fact that upwards of 50\% of 
the stars in any one globular cluster may be binaries and that the majority 
of these will be primordial (Hut et al. 1992). 
On the upside it has been shown by Makino \& Hut (1990) that the computational 
cost of primordial binaries becomes (relatively) less of a problem when 
$N$ is increased. 

The next generation machine, the GRAPE-6 (Makino 2001), became available
in early 2001 and has provided a further leap forward for $N$-body methods.
A single GRAPE-6 chip has an operating speed of $\sim 30\,$Gflops so that
a 32-chip board represents $1\,$Tflops of computing power.
The peak speed of GRAPE-6 is up to 100 times faster than GRAPE-4.
This exciting development means that models of $50\,000$ stars can be
simulated from formation to death, i.e. $\sim 10\,$Gyr,
in a week of wall-clock time.
As such, it is now possible to produce realistic $N$-body models of
small globular clusters for the first time.
The largest Galactic globular clusters are still out of reach for a single 
GRAPE-6 board.  
However, at the University of Tokyo, where the GRAPE project is based, 
work is underway to link many GRAPE-6 boards in parallel and simulate systems 
containing upwards of a hundred-thousand stars\footnote{We refer the interested 
reader to astrogrape.org for further information on the GRAPE project.}. 
To tackle the million body problem we will have to wait for the GRAPE-8 
which is expected to run at Petaflops speed. 
 
The Aarseth {\tt NBODY4} code (Aarseth 1999) has been developed to investigate 
the evolution of star clusters utilising the GRAPE hardware.  
Importantly {\tt NBODY4} incorporates algorithms for the detailed treatment 
of stellar evolution and the full range of possible interactions within 
binary stars (Hurley et al. 2001, and references within). 
In the dense environment of a star cluster it is possible for the
orbital parameters of a binary to be significantly perturbed owing
to close encounters with nearby stars.
As a result the orbit may even become chaotic (Mardling \& Aarseth 2001). 
Collisions may occur and binaries can form as a result of tidal capture 
(Fabian, Pringle \& Rees 1975). 
Gravitational encounters between single stars and binaries, or binary-binary 
encounters, can lead to an exchange interaction, where one star finds it 
energetically preferable to leave its current partner and become bound to 
another (Heggie 1975; Heggie \& Hut 2002). 
Other possible outcomes of these {\it scattering} incidents include the 
destruction of the one (or two) binary system(s) to leave only single 
stars or formation of a triple system. 
Rather than being instantaneous events, exchange interactions often involve 
a period of indecisiveness where stars swap between partners in quick 
succession before choosing a companion to remain with 
(see McMillan, Hut \& Makino 1991 for an early example of such behaviour). 
All these processes are accounted for in {\tt NBODY4}. 
This allows the generation and interaction of the full range of stellar 
populations possible. 
A similar effort is represented by the {\tt kira} code 
(Portegies Zwart et al. 2001b) which also takes advantage of the GRAPE hardware. 

\section{Promiscuous Stars}
\label{s:stars}

\subsection{Cluster Models} 

The use of $N$-body codes on the GRAPE hardware has already given rise 
to a number of important results regarding star cluster evolution. 
Aspects studied include the effect of dynamical evolution on the 
mass-functions of globular clusters (Vesperini \& Heggie 1997), the 
effect of the Galactic tidal field on the evolution of star clusters 
(Giersz \& Heggie 1997), and the evolution of cluster morphology 
(Boily, Clarke \& Murray 1999). 
The dependence of cluster lifetimes on their size and the tidal field in 
which they reside was recently investigated by Baumgardt (2001) who 
found that the lifetimes do {\it not} scale with the relaxation timescale, 
as had previously been assumed. 
Portegies Zwart et al (2001a) simulated the evolution and disruption 
of young compact star clusters near the Galactic centre and 
discussed the likelihood of their being observed. 

Hurley et al. (2001) demonstrated the effectiveness of the cluster 
environment in modifying the nature of the stars it contains by modelling 
the BS population of M67. 
Not only was the number of BSs produced twice that expected 
from standard binary evolution but formation paths for all the various 
types observed in M67 were created. 
The models of young Galactic open clusters performed by 
Portegies Zwart et al. (2001b) also exhibited strong dynamical activity. 
These studies have highlighted the importance of combining dynamics with 
stellar evolution. 

Within each of these open cluster size simulations there exist 
numerous interactions between stars that are often glossed over when 
discussing the simulation result. 
However, each of these interactions is interesting in its own right and 
we now take the opportunity to illustrate some of these fascinating tales 
in detail. 

\subsection{Lurid Examples} 

The first documentation of the behaviour of a particular star in 
a $N$-body simulation involving stellar evolution, and the motivation for 
this paper, was presented by Hurley et al. (2001). 
This star began life as a $1.33 M_\odot$ single star.  
Four exchange interactions and one tidal capture event later, involving 
a total of four collisions within eccentric binaries, its mass (after 
$4\,100\,$Myr of cluster evolution) had grown to $7.7 M_\odot$, an amazing 
factor of six greater than the MS turn-off mass at that time. 
At various intervening points in the evolution this star would have been 
observed as a single super-BS, a super-BS in a binary, a 
BS in an eccentric binary, or a BS in a triple system. 
{\it What we want to emphasise here is that such promiscuous behaviour 
is not at all rare.} 

\begin{figure}
\plotone{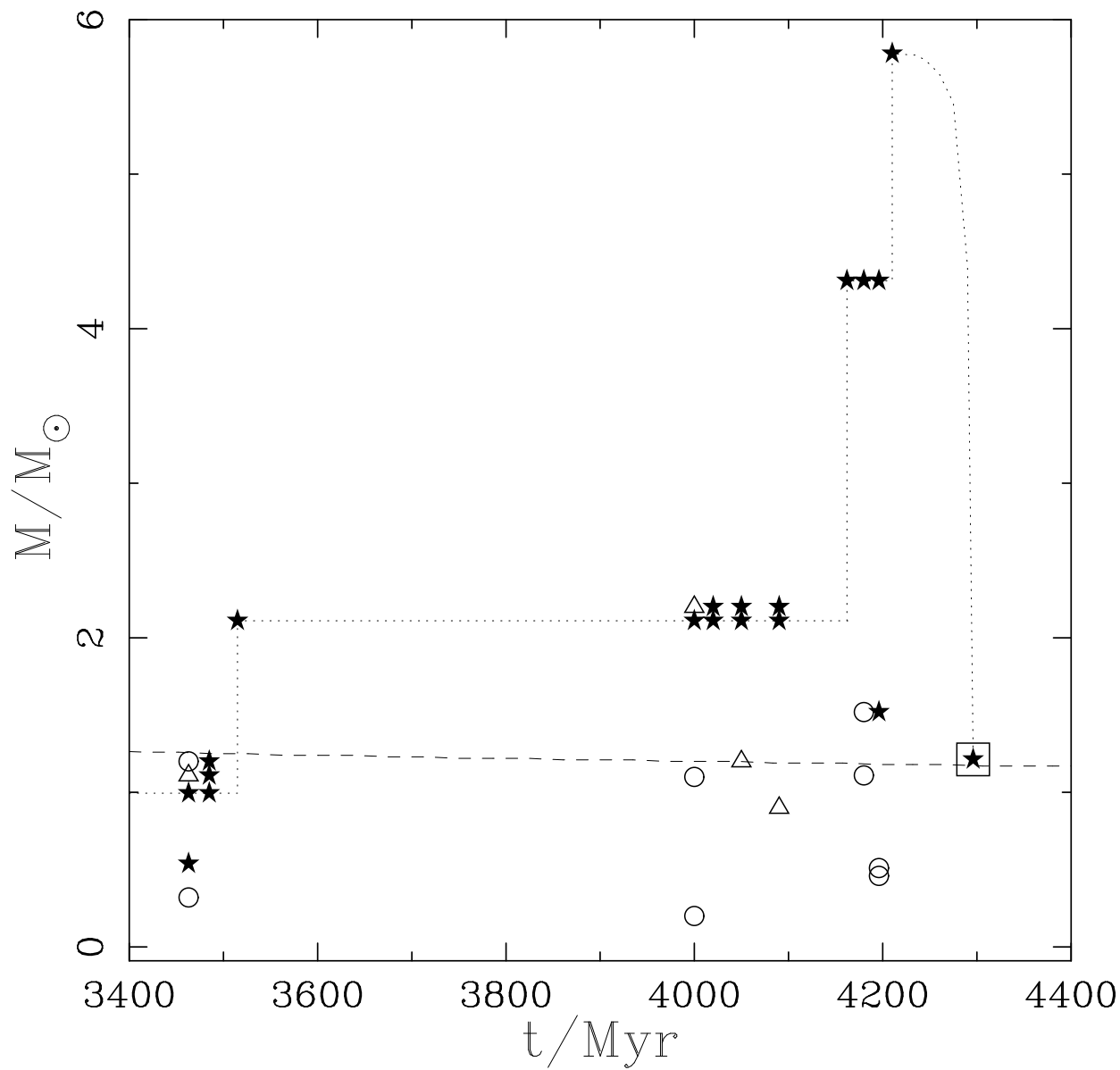}
\caption{
Mass as a function of time for the $0.99 M_\odot$ star that becomes a $5.78 M_\odot$ 
blue straggler, and the various stars and binaries that it interacts with along the way. 
The promiscuous star is shown as a solid star symbol, as is its companion in any 
binaries that it forms, and its path is traced out by the dotted line. 
Interacting single stars ($\bigtriangleup$) and binaries ($\circ$ for each component) 
are distinguished. 
The boxed symbol at $4\,296\,$Myr represents the $1.21 M_\odot$ oxygen-neon-WD that the 
$5.78 M_\odot$ blue straggler evolves to at that time (via a giant phase). 
The main-sequence turn-off mass is also shown (dash-dot line). 
\label{f:fig1}}
\end{figure}

Consider a similar case involving BS formation that arose in a recent 
simulation on GRAPE-6 aimed at modelling the evolution of planetary 
systems in star clusters (Hurley \& Shara 2002). 
At the beginning of the simulation the star of interest had a mass of 
$0.99 M_\odot$ and was in a primordial binary with a period of 
$P = 6\,513\,$yr and eccentricity $e = 0.73$. 
The companion star had a mass of $0.54 M_\odot$ and the binary was situated 
just beyond the cluster half-mass radius. 
The metallicity of the stars in this simulation was $Z = 0.004$. 
After $40\,$Myr the orbital parameters of this wide binary had been reduced 
to $P = 63\,$yr and $e = 0.13$ owing to perturbations resulting from 
weak gravitational encounters with nearby stars. 
At $T \simeq 490\,$Myr the binary had drifted inside the cluster half-mass 
radius and was involved in an exchange interaction with a $0.29 M_\odot$ 
star. 
Subsequently the low-mass interloper was ejected from the 3-body system 
leaving the original binary intact. 
Following this tumultuous but short-lived relationship, the binary 
partners remained quietly monogamous until becoming involved in a 5-body 
interaction at $T = 3\,463\,$Myr, by which time the binary had sunk  
further towards the cluster centre, through mass segregation, 
lying slightly outside of the core. 
The other participants in this encounter were a $1.12 M_\odot$ single star 
and a binary with component masses of $1.17$ and $0.32 M_\odot$ and 
$P = 58\,$yr. 
Initially the $0.54 M_\odot$ star was ejected, destroying the primordial 
binary, and a quasi-stable 4-body system remained. 
Shortly afterwards the $0.32 M_\odot$ star was also ejected and the 
$0.99 M_\odot$ and $1.12 M_\odot$ stars formed a binary with 
$P = 51\,$yr and $e = 0.95$ in a triple system with the $1.17 M_\odot$ star. 
The presence of the third body gradually drove the eccentricity of the inner 
binary up to 0.99 at which point ($T = 3\,515\,$Myr) the two MS stars collided  
to form a $2.11 M_\odot$ BS which did not remain bound to the third star. 
Figure~\ref{f:fig1} documents this interaction, and all subsequent interactions 
involving the initially $0.99 M_\odot$ star, in terms of the masses of the stars 
involved. 
We note that the $0.54 M_\odot$ star actually escaped from the cluster, with 
a velocity exceeding the stellar velocity dispersion by a factor of three,  
approximately $10\,$Myr after it was removed from its primordial companion 
- but as a direct result of the energy exchanged in that interaction. 

The BS (now significantly more massive than the average cluster star) 
quickly sank inside the core of the cluster where, at $T \simeq 4\,000\,$Myr, 
it was involved in an exchange interaction with a $2.2 M_\odot$ single star 
and a primordial binary. 
The single star, itself a BS formed via the coalescence of two MS 
stars in a semi-detached binary, formed a bound pair with the original BS. 
This BS-BS binary had a period of $487\,$yr and an eccentricity of 0.72. 
Its orbit was perturbed by interactions with a $1.2 M_\odot$ star at 
$T = 4\,049\,$Myr and with a $0.9 M_\odot$ star at $T = 4\,092\,$Myr which 
increased the eccentricity to 0.98. 
The orbit then became chaotic 
and at $T = 4\,162\,$Myr the BSs collided to form a $4.31 M_\odot$ super-BS. 
At $T = 4\,180\,$Myr the super-BS exchanged itself into a binary comprising 
a $1.11$ and a $1.52 M_\odot$ star and left bound to the more massive of the 
two, also a BS. 
Then, in a final showdown, this binary formed a 4-body system with another 
binary and at $T = 4\,210\,$Myr the $4.31 M_\odot$ super-BS and the 
$1.52 M_\odot$ BS collided to form a $5.78 M_\odot$ super-BS. 
Being as massive as it was, almost five times greater than the MS turn-off 
mass at the time, this promiscuous star quickly evolved off the MS and 
shortly afterwards lost its envelope on the asymptotic giant branch to 
become an oxygen-neon-WD. 

This lurid example, and the one presented by Hurley et al. (2001), highlight 
an important point. 
Though the possibility of any particular star becoming a BS as a result of 
a dynamical encounter is quite random, 
{\it once a star does become a BS and hence one of the most massive 
stars in the cluster, subsequent interactions are almost inevitable.} 
They also show that if a binary emerges from the indiscriminate and 
short-lived relationship that is an exchange interaction it 
will more often than not comprise the two most massive stars involved 
in the interaction, i.e. in this society it is desirable to be heavy. 

In a typical GRAPE-6 simulation with $N = 20\,000$, initially comprised of 
$18\,000$ single stars and $2\,000$ binaries, where the evolution was 
followed for $5\,$Gyr, the number of exchange interactions observed was 
$\sim 500$. 
These involved 730 different stars with some stars taking part in multiple 
interactions. 
The number of stars that swapped partner once was 494, twice was 105, 
three times was 48, four times was 27, and 14 stars swapped partner on 
five occasions. 
Amongst these were a number of re-marriages where a ``star changed its mind''  
and returned to its original partner. 
The component stars of one particularly flirtatious binary were actually 
involved in 22 exchange interactions, including 10 re-marriages. 

The total number of merger events observed during the entire 
simulation was 104. 
Of these, mass transfer in a primordial binary accounted for 46 cases while 
13 mergers came from mass transfer in a binary formed via an exchange 
interaction. 
The remainder were the result of collisions in eccentric binaries: 21 in 
primordial systems where the orbit was strongly perturbed by nearby stars.  

BS formation as a result of a hyperbolic collision is rare in simulations of 
open clusters but does occur. 
One example involved a $0.32 M_\odot$ star that began life in the core of 
the cluster and slowly drifted out to the half-mass radius, owing to 
mass-segregation, where at $1\,300\,$Myr it collided with a $0.93 M_\odot$ 
star. 
The relative velocity of the two stars at infinity was
$2.8 {\rm km} \, {\rm s}^{-1}$ and the collison product was assumed 
to be a fully-mixed $1.25 M_\odot$ MS star.  
When the cluster was $3\,850\,$Myr old this star first appeared as a
BS and by $4\,500\,$Myr, when the simulation ended, it had sunk 
inside the cluster core. 
The incidence of direct collisions will be greater in the higher density 
conditions of a globular cluster simulation. 
However, these are abrupt encounters and much less interesting than 
the sociable nature of exchange interactions and the binary systems 
they produce. 
The presence of binaries also acts to magnify the chance of collisions 
because in the case of a binary it is the semi-major axis that sets the 
relevant cross-section for collision, rather than the stellar radius which 
is used in the case of single stars. 
Here we have to be careful with the terminology used. 
If a binary is involved in a hyperbolic {\it interaction} any {\it collision} 
that results will occur in a two-step process and will not be {\it direct}: 
firstly a (resonant) capture may produce a hierarchical system and subsequently 
a collision can occur within this system. 

Interesting cases of stellar interactions are not limited to BS formation 
alone. 
Consider the case of the primordial binary comprised of MS stars of 
0.6 and $1.5 M_\odot$ with an eccentricity of 0.42 and an orbital 
period of $\sim 270\,$yr that was part of the same GRAPE-6 simulation 
mentioned above. 
Residing in the core of the cluster this binary suffered a series of weak
perturbations to its orbit which drove the eccentricity up to 0.95.
After the $1.5 M_\odot$ star became a sub-giant it was involved in two exchange
interactions, finally settling into a $33\,$yr orbit with eccentricity
of 0.9 about a $1.4 M_\odot$ MS star after $2\,130\,$Myr of evolution. 
The sub-giant then evolved onto the giant branch (GB) and by  
$T = 2\,160\,$Myr tidal forces within the binary had circularized the orbit 
resulting in a separation of $250 R_{\sun}$. 
While on the GB the $1.5 M_\odot$ star filled its Roche-lobe 
and a phase of common-envelope evolution began.
This stripped the envelope of the giant and left a helium-WD and a MS star
separated by $52 R_{\sun}$.
The MS star subsequently evolved to the GB, filled its Roche-lobe, and
another common-envelope event ensued, resulting in a pair of
0.4 and $0.3 M_{\sun}$ helium-WDs with an orbital period of $0.7\,$d.
Owing to gravitational radiation this system would easily merge within
$10^{10}\,$yr to possibly form a blue sub-dwarf star (Iben 1990), 
i.e. a helium-burning object with a thin hydrogen envelope. 

The final case that we choose to highlight involves the formation of 
a Thorne-\.{Z}ytkow object (TZO, Thorne \& \.{Z}ytkow 1977). 
Although TZOs have not been directly observed, they are thought 
to result from stellar mergers involving either a neutron star or a 
black hole, where the merger product is unstable and rapidly ejects 
all the material involved other than the neutron star or 
black hole, which remains. 
The system of interest began as a primordial binary with component 
masses $10.8$ and $5.3 M_\odot$, an eccentricity of 0.87 and a period 
of $1\,870\,$d. 
After $20\,$Myr of evolution the orbital period had been reduced to 
$1\,350\,$d owing to perturbations that had {\it hardened} this already 
{\it hard}\footnote{For a binary to be termed {\it hard} the 
magnitude of its binding energy must be greater than the mean kinetic 
energy of the cluster stars, otherwise it is {\it soft}. In stellar 
dynamics it is found that hard binaries will become harder, 
and soft binaries will be broken-up, as a result of encounters with 
other stars.} binary. 
A few Myr later the more massive of the two stars evolved onto the 
sub-giant branch and, as the region of convection within its envelope 
grew, tidal forces began to circularize the orbit.  
At $T = 23\,$Myr the primary star filled its Roche-lobe and began transferring 
mass to the companion. 
When the primary evolved onto the giant branch the rate of mass transfer 
accelerated so that the giant quickly overfilled the Roche-lobes of both 
stars to leave the $2.42 M_\odot$ helium core of the giant and the 
$5.31 M_\odot$ MS star contained within a common-envelope. 
Orbital friction then caused these two objects to spiral towards each other 
and the energy released was enough to drive off the envelope before they 
coalesced. 
This left a naked helium star and a MS star in a circular orbit with a 
$4.3\,$d period. 
The helium star then evolved onto the helium GB and filled its Roche-lobe. 
A period of stable mass transfer ensued until the envelope of the helium 
star had been removed and it became a $1.42 M_\odot$ WD of oxygen-neon 
composition. 
The orbital period at that stage was $10\,$d and the companion star was 
now a $6.07 M_\odot$ BS. 
At $T = 74\,$Myr the BS evolved off the MS onto the sub-giant branch and 
filled its Roche-lobe which lead to another common-envelope event and the 
formation of a naked helium star in an orbit of $P = 0.1\,$d with the 
oxygen-neon WD. 
When this helium star evolved onto the helium GB it also filled its 
Roche-lobe and began transferring mass to the WD. 
It was assumed that the helium-rich material would be accepted by the WD 
and steadily burnt on its surface. 
This quickly caused the mass of the WD to reach the critical Chandrasekhar 
mass, $1.44 M_\odot$, at which point it underwent an accretion-induced 
collapse to form a neutron star. 
The neutron star remained bound to the helium giant with an orbital period 
of $0.04\,$d. 
At $T = 94\,$Myr the helium giant again filled its Roche-lobe which lead to 
a third phase of common-envelope evolution and the formation of a TZO. 

In this particular instance dynamical interactions were not crucial to 
the outcome of the evolution: without perturbations to its orbit this 
binary would still have formed a TZO. 
However, we do see examples of similar systems being disrupted by 
encounters with other stars in the cluster so that the ultimate fate 
of the system is altered. 
Such occurences include the orbit being so strongly perturbed that the 
stars merge while both are on the MS. 
Alternatively a perturbation takes place just prior to the onset of a 
common-envelope phase, in a way that leads to coalescence of the two stars, 
rather than the formation of a close binary. 

The examples given here represent only a small fraction of the wealth 
of information regarding stellar interactions and populations that is 
generated by realistic $N$-body simulations of star clusters. 
Much of this information is quickly discarded if it is not of direct 
relevance to the initial purpose of that particular simulation. 
With GRAPE-6 increasing the $N$ that can be simulated in 
a reasonable timeframe there exists a distinct possibility that those 
working in the $N$-body field will rapidly be overwhelmed with data. 
Therefore, the creation of an archive for the results of $N$-body simulations 
that is accessable to all members of the scientific community is crucial 
(see Teuben et al. 2002 for a related discussion). 
This database will complement the HST stellar populations archive 
(Zurek: HST Proposal AR-9225) currently under construction at the 
American Museum of Natural History. 

It should also be noted that these interactions are not simply interesting: 
they have wide reaching implications. 
Take, for example, the case of type Ia supernovae which are important standard 
candles in cosmology and have recently been used to demonstrate that the 
expansion of the Universe is accelerating (Perlmutter et al. 1999). 
Short-period double-degenerate binaries in which the component WDs have a 
combined mass in excess of the critical Chandrasekhar mass and will merge 
within a Hubble time are thought to be possible progenitors of 
type Ia supernovae. 
Close inspection of the results of $N$-body simulations performed on GRAPE-6 
have shown that the production rate of these systems is remarkably 
enhanced - over an order of magnitude in open clusters and likely much 
more in globulars - relative to the field (Shara \& Hurley 2002).

\section{Summary}
\label{s:conclu}

The introduction of the GRAPE-6 hardware, coupled with $N$-body codes that
have the capability to produce realistic cluster models, places the
astrophysicist in the position of knowledgeable voyeur. 
An exciting aspect of this is the capability to investigate and
understand the range of stellar populations that are observed,
and how these are affected by dynamical interactions between cluster stars.
As we move to full utilization of the GRAPE-6, and the simulation of 
globular clusters, the fascinating tales of promiscuous stars in clusters 
will become very graphic indeed!

\acknowledgments

We are extremely grateful to Piet Hut and Jun Makino for their efforts in 
instigating and maintaining the GRAPE program. 
Immense gratitude is extended to Sverre Aarseth for his tireless work in
producing state-of-the-art $N$-body codes. 
We also thank Peter Eggleton, Rosemary Mardling, Steve McMillan, 
Seppo Mikkola, Onno Pols and Christopher Tout for important contributions 
made to the development of realistic $N$-body codes. 
Thanks to Simon Portegies Zwart and David Zurek for many stimulating 
discussions relating to the interactions of stars in clusters, and to 
the referee (Piet Hut) for a number of highly constructive suggestions. 
Doug Ellis' generous support has enabled AMNH to purchase new GRAPE-6
boards and we are most grateful to him.

\clearpage

\newpage 

\begin{thebibliography}{}
\bibitem[Aarseth (1963)]{aar63} Aarseth, S.J. 1963, \mnras, 126, 223 
\bibitem[Aarseth (1999)]{aar99} Aarseth, S.J. 1999, \pasp, 111, 1333 
\bibitem[Bailyn (1995)]{bai95} Bailyn, C.D. 1995, \araa, 33, 133 
\bibitem[Baumgardt (2001)]{bau01} Baumgardt, H. 2001, \mnras, 325, 1323 
\bibitem[Boily, Clarke \& Murray (1999)]{boi99} Boily, C.M., Clarke, C.J., 
    \& Murray, S.D. 1999, \mnras, 302, 399 
\bibitem[D'Amico et al. (2001)]{dam01} D'Amico, N., Lyne, A.G., 
    Manchester, R.N., Possenti, A., \& Camilo, F. 2001, \apj, 548, L171 
\bibitem[Fabian, Pringle \& Rees (1975)]{fab75} Fabian, A.C., Pringle, J.E., 
    \& Rees, M.J. 1975, \mnras, 172, 15 
\bibitem[Ferraro et al. (1999)]{fer99} Ferraro, F.R., Paltrinieri, B., 
    Rood, R.T., \& Dorman, B. 1999, \apj, 522, 983 
\bibitem[Giersz (2001)]{gie01} Giersz, M. 2001, \mnras, 324, 218 
\bibitem[Giersz \& Heggie (1997)]{gie97} Giersz, M., \& Heggie, D.C. 
    1997, \mnras, 286, 709  
\bibitem[Grindlay et al. (2001)]{gri01} Grindlay, J.E., Heinke, C., 
    Edmonds, P.D., \& Murray, S.S. 2001, Science, 292, 2290 
\bibitem[Guhathakurta et al. (1998)]{guh98} Guhathakurta, P., Webster, Z.T., 
    Yanny, B., Schneider, D.P., \& Bahcall, J.N. 1998, \aj, 116, 1757 
\bibitem[Heggie (1975)]{heg75} Heggie, D.C. 1975, \mnras, 173, 729 
\bibitem[Heggie \& Aarseth (1992)]{heg92} Heggie, D.C., \& Aarseth, S.J. 
    1992, \mnras, 257, 513 
\bibitem[Heggie \& Hut (2002)]{heg02} Heggie, D.C., \& Hut, P. 2002, 
    The Gravitational Million-Body Problem (Cambridge University Press), in press 
\bibitem[H\'{e}non (1961)]{hen61} H\'{e}non, M. 1961, Ann. d'Astrophys., 24, 369 
\bibitem[Hurley et al. (2001)]{hur01} Hurley, J. R., Tout, C. A., 
    Aarseth, S. J., \& Pols, O.R. 2001, \mnras, 323, 630
\bibitem[Hurley \& Shara (2002)]{hur02} Hurley, J. R., \& Shara, M.M. 2002, 
    \apj, in press (astro-ph/0108350) 
\bibitem[Hut et al. (1992)]{hut92} Hut, P., McMillan, S., Goodman, J., 
    Mateo, M., Phinney, E.S., Pryor, C., Richer, H.B., Verbunt, F., 
    \& Weinberg, M. 1992, \pasp, 104, 981 
\bibitem[Iben (1990)]{ibe90} Iben, I.Jr. 1990, \apj, 353, 215 
\bibitem[Joshi, Nave \& Rasio (2001)]{jos01} Joshi, K.J., Nave, C.P., 
    \& Rasio, F.A. 2001, \apj, 550, 691 
\bibitem[Leonard (1996)]{leo96} Leonard, P.J.T. 1996, ApJ, 470, 521 
\bibitem[Makino \& Hut (1990)]{mak90} Makino, J., \& Hut, P. 1990, \apj, 
    365, 208 
\bibitem[Makino \& Taiji (1998)]{mak98} Makino, J., \& Taiji, M. 1998, 
    Scientific Simulations with Special-Purpose Computers -- the GRAPE Systems 
    (New York: Wiley)  
\bibitem[Makino (2001)]{mak01} Makino, J. 2001, in ASP Conf. Ser. XX, 
    Stellar Collisions, ed. M. M. Shara (San Francisco: ASP), in press 
\bibitem[Mardling \& Aarseth (2001)]{mar01} Mardling, R.A.,
   \& Aarseth, S.J. 2001, \mnras, 321, 398 
\bibitem[McMillan (1986)]{mcm86} McMillan, S.L.W. 1996, in 
    The Use of Supercomputers in Stellar Dynamics, ed. 
    P. Hut \& S.L.W. McMillan (Springer-Verlag: Berlin), 156 
\bibitem[McMillan, Hut \& Makino (1991)]{mcm91} McMillan, S.L.W., Hut, P., 
    \& Makino, J. 1991, \apj, 372, 111
\bibitem[Meylan \& Heggie (1997)]{mey97} Meylan, G., \& Heggie, D. 
    1997, \aapr, 8, 1  
\bibitem[Mikkola \& Aarseth (1998)]{mik98} Mikkola, S., \& Aarseth, S.J. 
    1998, New Astronomy, 3, 309 
\bibitem[Moehler, Heber \& Durell (1997)]{moe97} Moehler, S., Heber, U., 
    \& Durell, P.R. 1997, \aap, 317, 83 
\bibitem[Perlmutter et al. (1999)]{per99} Perlmutter, S., et al. 1999,
   \apj, 517, 565
\bibitem[Portegies Zwart et al. (2001a)]{por01a} Portegies Zwart, S.F., 
    Makino, J., McMillan, S.L.W., \& Hut, P. 2001a, \apj, 546, L101  
\bibitem[Portegies Zwart et al. (2001b)]{por01b} Portegies Zwart, S.F., 
    McMillan, S.L.W., Hut, P., \& Makino, J. 2001b, \mnras, 321, 199 
\bibitem[Shara et al. (1996)]{sha96} Shara, M.M., Bergeron, L.E., 
    Gilliland, R.L., Saha, A., \& Petro, L. 1996, \apj, 471, 804 
\bibitem[Shara \& Hurley (2002)]{sha02} Shara, M.M., \& Hurley, J. R. 2002, 
    \apj, submitted 
\bibitem[Takahashi \& Portegies Zwart (2000)]{tak00} Takahashi, K., 
    \& Portegies Zwart, S.F. 2000, \apj, 535, 759  
\bibitem[Teuben et al. (2002)]{teu02} Teuben, P., DeYoung, D., Hut, P., 
    Levy, S., Makino, J., McMillan, S., Portegies Zwart, S., \& Slavin, S. 
    2002, Theory in a Virtual Observatory, 
    to appear in the proceedings of the ninth ADASS conference 
    (astro-ph/0111478) 
\bibitem[Thorne \& \.{Z}ytkow (1977)]{tho77} Thorne K.S., \& \.{Z}ytkow A.N. 
    1977, \apj, 212, 832
\bibitem[van den Berg et al. (2001)]{van01} van den Berg, M., Orosz, J., 
    Verbunt, F., \& Stassun, K. 2001, \aap, 375, 375 
\bibitem[Vesperini \& Heggie (1997)]{ves97} Vesperini, E., \& Heggie, D.C. 
    1997, \mnras, 289, 898 
\end{thebibliography}
\end{document}